\documentclass[12pt]{article}
\usepackage[super,compress]{cite}
\usepackage[super]{cite}
\usepackage{amsmath,amsfonts,euscript,amssymb}
\usepackage{epsfig}

\DeclareSymbolFont{bfitletters}{OML}{cmm}{bx}{it}
\DeclareSymbolFont{bfitoperators}   {OT1}{cmr} {m}{n}
\DeclareMathSymbol{\bfitomega}{\mathord}{bfitletters}{"21}
\DeclareMathSymbol{\bfitrho}{\mathord}{bfitletters}{"1A}
\DeclareMathSymbol{\bfitgamma}{\mathord}{bfitletters}{"0D}
\DeclareMathSymbol{\bfitchi}{\mathord}{bfitletters}{"1F}
\DeclareMathSymbol{\bfitxi}{\mathord}{bfitletters}{"18}
\DeclareMathSymbol{\bfitpi}{\mathord}{letters}{"19}

\newcommand{\be}{\begin{equation}}
\newcommand{\ee}{\end{equation}}
\newcommand{\bea}{\begin{eqnarray}}
\newcommand{\eea}{\end{eqnarray}}

\begin{document}

\begin{center}
{\Large\bf Hamiltonian equations of reduced conformal geometrodynamics in extrinsic time}\\
\vskip 1cm
{\bf Alexander E. Pavlov}
\vskip 0.5cm
{\it Institute of Mechanics and Power Engineering,
  Russian State Agrarian University ---
  Moscow Timiryazev Agricultural Academy, Moscow, 127550, Russia}
\end{center}




%

\vskip 2cm
\begin{abstract}
The reduced vacuum Hamiltonian equations of conformal geometrodynamics of compact manifolds in extrinsic time are written. This is achieved by generalizing the theorem of implicit function derivative to the functional analysis.

\end{abstract}

\newpage
\section{Introduction}

The Hamiltonian dynamics of gravitational field is commonly formulated
in redundant variables in an extended functional phase space as a consequence of covariant
description of Einstein's theory. A time parameter should be conjugated to the Hamiltonian
constraint. The Hamiltonian formulation of the theory makes it possible to reveal
the physical meaning of geometrical variables. The problem of writing of the vacuum Einstein's equations in unconstrained variables for the compact cosmological models is actual. The reduced phase space is the cotangent bundle
of the Teichm${\rm\ddot{u}}$ller space of conformal structures on compact spacelike hypersurfaces \cite{Fischer:1996qg}. The Hamiltonian is the volume functional of these hypersurfaces. Hamiltonian dynamics is constructed in York's time~\cite{York}. The problem is that the Hamiltonian density as volume measure  is not expressed in explicit form from the Hamiltonian constraint (Lichnerowicz--York elliptic differential equation).
This makes it difficult to obtain a Hamiltonian flow.
In the present paper we obtain the Hamiltonian equations of motion. This is achieved by
generalizing the theorem of implicit function derivative theorem from the mathematical analysis to the functional analysis.

Let the spacetime ${\mathcal M}={\mathbb{R}}^1\times\Sigma_t$
is foliated into a family of space-like hypersurfaces $\Sigma_t$, labeled
by the time coordinate $t$ with just three spatial coordinates on each
slice $(x^1, x^2, x^3)$.
The first quadratic form
\be\label{spacemetric}
{\bfitgamma}:=\gamma_{ik}(t, {\bf x}){\rm\bf d}x^i\otimes {\rm\bf d}x^k
\ee
defines the induced metric on every slice $\Sigma_t$.
The Hamiltonian dynamics is built of the ADM - variational functional
\be\label{ADMvar}
S_{\rm ADM}=\int\limits_{t_I}^{t_0}
\, {\rm d}t\int\limits_{\Sigma_t} {\rm d}^3x\left(\pi^{ij}\frac{\partial\gamma_{ij}}{\partial t}-
N{\cal H}_\bot-N^i{\cal H}_i\right),
\ee
where ADM units: $c=1, 16\pi G=1$ were used. The variation of the action (\ref{ADMvar}) by the lapse function $N$ leads to
the Hamiltonian constraint expressed via components of the extrinsic curvature $K_{ij}$ or by the components of the momentum densities $\pi^{ij}$:
\bea\nonumber
{\cal H}_\bot&=&\sqrt\gamma\left(K_{ij}K^{ij}-K^2-R\right)\\
&=&\frac{1}{2\sqrt\gamma}\left(
\gamma_{ik}\gamma_{jl}+\gamma_{il}\gamma_{jk}-\gamma_{ij}\gamma_{kl}\right)
\pi^{ij}\pi^{kl}-\sqrt\gamma R.\label{Hc}
\eea
Here, $\gamma$ is the determinant of the metric tensor, $K$ is the trace of the extrinsic curvature tensor $K:=K_{ij}\gamma^{ij}$, and $R$ is the Ricci scalar curvature.
Varying the action (\ref{ADMvar}) by the shift functions $N^i$, we get the momentum constraints:
\be\label{mc}
{\cal H}_i=2\sqrt\gamma\left(\nabla_j K^j_i-\nabla_i K\right)=-2\nabla_j\pi^j_i,
\ee
where the connection $\nabla_i$ is associated with the metric $\gamma_{ij}$.

The variations of the action (\ref{ADMvar}) by the canonical variables $\pi^{kl}(t,{\bf x})$ and
$\gamma_{ij}(t,{\bf x})$ lead to the kinematical equations:
\bea
\frac{\delta}{\delta\pi^{ij}}S_{\rm ADM}=
\frac{\partial\gamma_{ij}}{\partial t}
&=&-2N K_{ij}+\nabla_i N_j+\nabla_j N_i\nonumber\\
&=&\frac{2N}{\sqrt\gamma}\left(\pi_{ij}-\frac{1}{2}\gamma_{ij}\pi\right)+\nabla_i N_j+\nabla_j N_i,\label{eqpi}
\eea
and to the dynamical equations
\bea
\frac{\delta}{\delta\gamma_{ij}}S_{\rm ADM}&=&-\frac{\partial\pi^{ij}}{\partial t}=
N\sqrt{\gamma}\left(R^{ij}-\frac{1}{2}\gamma^{ij}R\right)-
\frac{N}{2\sqrt\gamma}\gamma^{ij}\left(\pi^{mn}\pi_{mn}-\frac{1}{2}\pi^2\right)\nonumber\\
&+&\frac{2N}{\sqrt\gamma}\left(\pi^{im}\pi^j_m-\frac{1}{2}\pi\pi^{ij}\right)-\sqrt\gamma\left(\nabla^i\nabla^j N-\gamma^{ij}\triangle N\right)\nonumber\\
&-&\nabla_m\left(\pi^{ij} N^m\right)+\nabla_m N^i\pi^{mj}+\nabla_m N^j\pi^{mi},\label{eqg}
\eea
where $\triangle:=\nabla_i\nabla^i$ is the Laplacian.

\section{Conformal Decomposition}

The equations of motion (\ref{eqpi}), (\ref{eqg}) contain the unknown Lagrange multipliers. To obtain the dynamical variables the conformal transformation are implemented~\cite{T}
\begin{equation}\label{Psifactor}
\gamma_{ij}:=\phi^4\tilde\gamma_{ij},\qquad \phi^4:=\sqrt[3]{\gamma}.
\end{equation}
To the conformal variables
\begin{equation}\label{generalized}
{{\tilde\gamma_{ij}:=\frac{\gamma_{ij}}{\sqrt[3]{\gamma}},\qquad
\tilde\pi^{ij}:=\sqrt[3]{\gamma}\left(\pi^{ij}-\frac{1}{3}\pi\gamma^{ij}\right)}},
\end{equation}
where $\pi:=\gamma_{ij}\pi^{ij}$, we add the canonical pair:
\begin{equation}\label{DiracTpi}
T:=\frac{2}{3}\frac{\pi}{\sqrt{\gamma}}=\frac{4}{3}K,\qquad {\cal H}=\sqrt\gamma.
\end{equation}
Formulae~(\ref{generalized}) and~(\ref{DiracTpi}) define the Dirac's mapping.
The Lie--Poisson structure of the new variables in the extended phase space
$\Gamma_T [\tilde\gamma_{ij}, \tilde\pi^{ij}; T, {\cal H}]$ is the following
\begin{eqnarray}
 \{T(t,{\bf x}), {\cal H}(t,{\bf x}')\}&=&-\delta ({\bf x}-{\bf x}'),
\label{PoissonD} \\
\{\tilde\gamma_{ij}(t,{\bf x}),\tilde\pi^{kl}(t,{\bf x}')\}&=&
(\delta_i^k\delta_j^l+\delta_i^l\delta_j^k
-\frac{1}{3}\tilde\gamma^{kl}\tilde\gamma_{ij})
\delta ({\bf x}-{\bf x}'),
\label{PoissonG} \\
 \{\tilde\pi^{ij}(t,{\bf x}),\tilde\pi^{kl}(t,{\bf x}')\}&=&
\frac{1}{3}(\tilde\gamma^{kl}\tilde\pi^{ij}
-\tilde\gamma^{ij}\tilde\pi^{kl}) \delta ({\bf x}-{\bf x}'). \label{PoissonPi}
\end{eqnarray}
The subalgebra (\ref{PoissonD}) of the canonical pair $(T,{\cal H})$
is split out of the algebra (\ref{PoissonG}), (\ref{PoissonPi}).

Then we extract the traceless part $A^{ij}$ from the extrinsic curvature tensor
\be\nonumber
K^{ij}=A^{ij}+\frac{1}{3}K\gamma^{ij}.
\ee
Under the conformal transformation (\ref{Psifactor}) its components transform according to
\be\nonumber
A^{ij}=\phi^{-10}\tilde{A}^{ij},\qquad A_{ij}=\phi^{-2}\tilde{A}_{ij}.
\ee
They are connected with the components  of the conformal momentum densities
\be\nonumber
\tilde{A}^{ij}=-\tilde\pi^{ij},\qquad \tilde{A}_{ij}=-\tilde\pi_{ij}.
\ee
The conformal Ricci scalar $\tilde{R}$ is expressed from the Ricci scalar $R$:
\be\nonumber
R=\phi^{-4}\tilde{R}-8\phi^{-5}\tilde{\triangle}\phi,
\ee
where $\tilde\triangle:=\tilde\nabla_i\tilde\nabla^i$ is the conformal Laplacian, and $\tilde\nabla_i$ is the conformal connection associated with the conformal metric $\tilde\gamma_{ij}$.
The conformal Hamiltonian constraint (\ref{Hc})
\begin{equation}\label{confH}
\tilde{\cal H}_\bot=
\tilde\pi_{ij}\tilde\pi^{ij}\phi^{-6}+8\phi\tilde\Delta\phi-\tilde{R}\phi^2-\frac{3}{8}T^2\phi^6,
\end{equation}
and the conformal momentum constraints (\ref{mc}):
\be\label{confmom}
\tilde{\cal H}^i=-2\phi^{-4}\tilde\nabla_j\tilde\pi^{ij}-\frac{4}{3}\phi^2\tilde\nabla^i K.
\ee

York proposed the constant curvature condition (CMC) condition~\cite{York}: $T=t$ to fix the spacetime slicing.
This gauge allowed decompose the conformal momentum densities into longitudinal $\tilde\pi^{ij}_L$ and traceless-transverse $\tilde\pi^{ij}_{TT}$ parts:
\be\nonumber
\tilde\pi^{ij}:=\tilde\pi^{ij}_L+\tilde\pi^{ij}_{TT}.
\ee
The longitudinal part $\tilde\pi^{ij}_L$ is the constrained part and obtained as a solution of the linear elliptic differential equations (\ref{confmom}).

\section{Conformal Hamiltonian Equations of Motion}

Expressing the conformal factor $\phi$ via the Hamiltonian density $\phi={\cal H}^{1/6}$, we substitute it into the Hamiltonian constraint (\ref{confH}):
\begin{equation}\label{confHamconstr}
\tilde{\cal H}_\bot=
\frac{1}{2}\left(\tilde\gamma_{ik}\tilde\gamma_{jl}+\tilde\gamma_{il}\tilde\gamma_{jk}\right)\tilde\pi^{ij}\tilde\pi^{kl}
{\cal H}^{-1}+8{\cal H}^{1/6}\tilde\Delta {\cal H}^{1/6}-\tilde{R}{\cal H}^{1/3}-\frac{3}{8}T^2{\cal H}.
\end{equation}
The reduced ADM action (\ref{ADMvar}) then reads
\be\label{Sred}
S_{\rm reduced}=\int\limits_{T_I}^{T_0}
\, {\rm d}T\int\limits_{\Sigma_T} {\rm d}^3x\left(\tilde\pi^{ij}\frac{\partial\tilde\gamma_{ij}}{\partial T}-
{\cal H} [\tilde\pi^{ij}, \tilde\gamma_{ij}; T]-N^i \tilde{\cal H}_i[\tilde\pi^{ij}, \tilde\gamma_{ij}]\right).
\ee
The conformal Hamiltonian density ${\cal H}$ is a functional of the variables $\tilde\pi^{ij},$ $\tilde\gamma_{ij}$
and a function of the time T; $\tilde{\cal H}_i$ are the generators of changing of the coordinates in the hypersurface. The Hamiltonian
\be\label{Hamiltonian}
H:=\int\limits_{\Sigma_T}{\rm d}^3x\, {\cal H}[\tilde\pi^{ij},\tilde\gamma_{ij}; T]
\ee
generates the dynamics of the gravitational field. Unfortunately, we have not its explicit form.
Below, we can find the derivatives of the Hamiltonian (\ref{Hamiltonian}) with respect to conformal variables.

The variation of the functional of the conformal Hamiltonian constraint (\ref{confHamconstr})
\be\label{functionalH}
\tilde{H}_\bot:=\int\limits_{\Sigma_t}{\rm d}^3x\, \tilde{\cal H}_\bot [\tilde\pi^{ij},\tilde\gamma_{ij}; {\cal H}, T]
\ee
on a slice $T$ is zero:
\begin{equation}\label{dtildeH}
\delta\tilde{H}_\bot=
\int\limits_{\Sigma_t} {\rm d}^3x\left(
\frac{\delta \tilde{H}_\bot}{\delta {\cal H}}\delta {\cal H}+
\frac{\delta \tilde{H}_\bot}{\delta\tilde\pi^{ij}}\delta {\tilde\pi^{ij}}+
\frac{\delta \tilde{H}_\bot}{\delta\tilde\gamma_{ij}}\delta {\tilde\gamma_{ij}}\right)=0.
\end{equation}
The variation of the Hamiltonian density can be presented as
\begin{equation}\nonumber
\delta {\cal H}=\frac{\partial {\cal H}}{\partial\tilde\pi^{ij}}\delta\tilde\pi^{ij}+
\frac{\partial {\cal H}}{\partial\tilde\gamma_{ij}}\delta\tilde\gamma_{ij}.
\end{equation}
After substitution $\delta\cal{H}$ into (\ref{dtildeH}) one gets
\begin{equation}
\delta \tilde{H}_\bot=\int\limits_{\Sigma_t} {\rm d}^3x\left[
\left(
\frac{\delta \tilde{H}_\bot}{\delta {\cal H}}\frac{\partial {\cal H}}{\partial {\tilde\pi^{ij}}}+
\frac{\delta \tilde{H}_\bot}{\delta\tilde\pi^{ij}}\right)\delta\tilde\pi^{ij}+
\left(\frac{\delta \tilde{H}_\bot}{\delta {\cal H}}\frac{\partial {\cal H}}{\partial\tilde\gamma_{ij}}+
\frac{\delta \tilde{H}_\bot}{\delta\tilde\gamma_{ij}}
\right)\delta\tilde\gamma_{ij}
\right]=0.
\end{equation}
Taking into account the independence of the variations, we obtain the derivatives
\begin{equation}\label{der}
\frac{\partial {\cal H}}{\partial\tilde\pi^{ij}}=-\frac{\delta \tilde{H}_\bot /\delta\tilde\pi^{ij}}
{\delta \tilde{H}_\bot /\delta {\cal H}},\qquad
\frac{\partial {\cal H}}{\partial\tilde\gamma_{ij}}=-\frac{\delta \tilde{H}_\bot /\delta\tilde\gamma_{ij}}
{\delta \tilde{H}_\bot /\delta {\cal H}}.
\end{equation}

The Hamiltonian $H$ generates the phase flow in the phase space
$\Gamma [\tilde\gamma_{ij}, \tilde\pi^{ij}]$ on the Poisson brackets (\ref{PoissonG}), (\ref{PoissonPi}):
\begin{equation}\label{ddTg}
\frac{\partial}{\partial T}\tilde\gamma_{ij}(x)=\{\tilde\gamma_{ij}(x),H\}=\int\limits_{\Sigma_t}{\rm d}^3x'\,\{\tilde\gamma_{ij}(x),\tilde\pi^{kl}(x')\}
\frac{\partial {\cal H}}{\partial\tilde\pi^{kl}}(x'),
\end{equation}
\begin{eqnarray}\nonumber
\frac{\partial}{\partial T}\tilde\pi^{ij}(x)=\{\tilde\pi^{ij}(x),H\}&=&\int\limits_{\Sigma_t}{\rm d}^3x'\,\{\tilde\pi^{ij}(x),\tilde\pi^{kl}(x')\}
\frac{\partial {\cal H}}{\partial\tilde\pi^{kl}}(x')\\
&+&\int\limits_{\Sigma_t}{\rm d}^3x'\,\{\tilde\pi^{ij}(x),\tilde\gamma_{kl}(x')\}
\frac{\partial {\cal H}}{\partial\tilde\gamma_{kl}}(x').\label{ddTpi}
\end{eqnarray}
Let us calculate the functional derivative of (\ref{functionalH}) with respect to Hamiltonian density:
\begin{eqnarray}\nonumber
\frac{\delta \tilde{H}_\bot}{\delta {\cal H}(x)}=&-&\frac{1}{2}(\tilde\gamma_{ik}\tilde\gamma_{jl}+\tilde\gamma_{il}\tilde\gamma_{jk})\tilde\pi^{ij}\tilde\pi^{kl}
{\cal H}^{-2}(x)-\frac{1}{3}\tilde{R}{\cal H}^{-2/3}(x)-\frac{3}{8}T^2\\
&+&\frac{8}{3}{\cal H}^{-5/6}(x)\tilde\Delta {\cal H}^{1/6}(x).\label{dHam}
\end{eqnarray}
The functional derivatives of (\ref{functionalH}) with respect to the conformal momentum densities:
\begin{equation}
\frac{\delta \tilde{H}_\bot}{\delta\tilde\pi^{ij}(x)}=(\tilde\gamma_{ik}\tilde\gamma_{jl}+
\tilde\gamma_{il}\tilde\gamma_{jk})\tilde\pi^{kl}{\cal H}^{-1}(x),\label{dpi}
\end{equation}
and with respect to the conformal metric:
\begin{equation}
\frac{\delta \tilde{H}_\bot}{\delta\tilde\gamma_{ij}(x)}=2\tilde\gamma_{kl}\tilde\pi^{ik}\tilde\pi^{jl}{\cal H}^{-1}-
8\tilde\nabla^i {\cal H}^{1/6}\tilde\nabla^j {\cal H}^{1/6}-
\tilde{R}^{ij}{\cal H}^{1/3}.\label{dgamma}
\end{equation}
Substituting the functional derivatives (\ref{dHam}), (\ref{dpi}), (\ref{dgamma}) into (\ref{der}), we get the partial derivatives. Then, taking into account the algebra (\ref{PoissonG}), (\ref{PoissonPi}), we substitute these partial derivatives into (\ref{dHam}), (\ref{dpi}) and obtain the Hamiltonian equations of motion of gravitational field.
\be\label{gravgamma}
\frac{\partial}{\partial T}\tilde\gamma_{ij}(x)=-\frac{4/{\cal H}}{\delta \tilde{H}_\bot /\delta H}
\tilde\pi_{ij}(x).
\ee
They present the kinematical equations for the conformal variables as (\ref{eqpi}). The next are the dynamical equations as (\ref{eqg}):
\bea\nonumber
&&\frac{\partial}{\partial T}\tilde\pi^{ij}(x)=
\frac{4/{\cal H}}{\delta \tilde{H}_\bot /\delta {\cal H}}\tilde\gamma_{kl}\tilde\pi^{ik}\tilde\pi^{jl}\\
&-&\frac{8}{\delta \tilde{H}_\bot /\delta {\cal H}}
\left(
2\left(\tilde\nabla^i{\cal H}^{1/6}\right)\left(\tilde\nabla^j{\cal H}^{1/6}\right)
-\frac{1}{3}\tilde\gamma^{ij}
\left(\tilde\nabla_k{\cal H}^{1/6}\right)\left(\tilde\nabla^k{\cal H}^{1/6}\right)
\right)\nonumber\\
&-&\frac{{\cal H}^{1/3}}{\delta \tilde{H}_\bot /\delta {\cal H}}\left(2\tilde{R}^{ij}-\frac{1}{3}\tilde{R}\tilde\gamma^{ij}\right).\label{gravpi}
\eea

In the present paper we did not make simplifications anywhere, so the form of the equations looks rather complicated.
Their advantage in comparison with the ADM equations (\ref{eqpi}), (\ref{eqg})
is that they do not contain Lagrange multipliers.
They can be useful under considering model problems and perturbation theory, since their appearance will be simplified.

\section{Conclusion and Discussion}

The Ricci flow of three-manifolds was studied in Ref. 3, and the conformal Ricci flow in Ref. 4. that says about the importance of the subject. In general case, the Hamiltonian constraint is an elliptic differential equation for the Hamiltonian density. For the systems with finite degrees of freedom it becomes an algebraic equation.
For a homogeneous and isotropic minisuperspace model the reduction was undertaken in Ref. 5, and for an anisotropic model in Ref. 6. The York's time is proportional to the Hubble parameter.
The global intrinsic time was constructed in Ref. 7. It was achieved by averaging of geometric characteristics by hypersurfaces of constant coordinate time.
The Hamiltonian equations of motion in the intrinsic time are written in Ref. 8.
In this case, the Hamiltonian and time change places.
The advantage of this approach is that one can express the Hamiltonian from the Hamiltonian constraint explicitly.
But it will not be possible to split off the longitudinal components of the momentum densities.


\end{document}